\documentclass[showpacs,floatfix,eqsecnum,aps,nofootinbib, amssymb]{revtex4}

\usepackage{longtable}
\usepackage{graphicx}
\usepackage{lscape}

\begin{document}

\title{Kink-induced symmetry breaking patterns in brane-world SU(3)$^3$ trinification models.}  
\author{Alison Demaria}\email{a.demaria@physics.unimelb.edu.au} \affiliation{School
of Physics, Research Centre for High Energy Physics, The University of
Melbourne, Victoria 3010, Australia} \author{Raymond
R. Volkas}\email{r.volkas@physics.unimelb.edu.au} \affiliation{School
of Physics, Research Centre for High Energy Physics, The University of
Melbourne, Victoria 3010, Australia}

\begin{abstract}
The trinification grand unified theory (GUT)
has gauge group $SU(3)^3$ and a discrete symmetry permuting the $SU(3)$
factors. In common with other GUTs, the attractive nature of the
fermionic multiplet assignments is obviated by the complicated
multi-parameter Higgs potential apparently
needed for phenomenological reasons, and also by vacuum expectation value (VEV)
hierarchies {\it within} a given multiplet. 
This motivates the rigorous
consideration of Higgs potentials, symmetry breaking patterns and
alternative symmetry breaking mechanisms in models with this gauge group.
Specifically, we study the recently proposed ``clash of symmetries''
brane-world mechanism to see if it can help with the symmetry breaking
conundrum. This requires a detailed analysis of Higgs potential global
minima and kink or domain wall solutions interpolating between the
disconnected global minima created through spontaneous discrete
symmetry breaking. Sufficiently long-lived metastable
kinks can also be considered.  We develop what we think is an
interesting, albeit speculative, brane-world scheme whereby the
hierarchical symmetry breaking cascade, trinification to left-right
symmetry to the standard model to colour cross electromagnetism,
may be induced without an initial hierarchy in vacuum expectation
values. Another motivation for this paper is simply to continue the
exploration of the rich class of kinks arising in models that are
invariant under both discrete and continuous symmetries.

\end{abstract}

\maketitle

\section{Introduction}

Symmetry principles are basic to quantum field theoretic models of
particle physics and to general relativity. In particle physics, the
fundamental fields are placed into representations of a gauge group
and then an invariant Lagrangian is constructed from the fields and
their 4-gradients. In some models, global symmetries are also imposed 
to either forbid unwanted terms or
to relate otherwise independent parameters. The standard model (SM)
Lagrangian is defined as the most general linear combination of all
$SU(3)_c \otimes SU(2)_L \otimes U(1)_Y$ gauge-invariant
renormalisable terms constructed from the familiar quark, lepton,
gauge and Higgs boson fields. The Higgs doublet self-interactions
cause the ground or vacuum state of the SM to respect the smaller
gauge group $SU(3)_c \otimes U(1)_Q$, where $Q$ is electric
charge. This electroweak spontaneous symmetry breaking (SSB) gives
masses to the $W$ and $Z$ bosons and the quarks and leptons. One of the
theoretically unsatisfactory features of the SM arises from the Higgs sector:
the proliferation of Yukawa coupling constants and hence
the {\it a priori} arbitrary nature of the fermion masses. The SM
Higgs potential, on the other hand, is rather appealing: it is very
simple, having only two free parameters, and it is impossible to
spontaneously break electromagnetic and colour gauge invariance.

Much effort has been expended on SM extensions where the
new gauge groups are larger than $SU(3)_c \otimes SU(2)_L \otimes
U(1)_Y$, the goal being to construct more predictive models while
retaining the phenomenological successes of the SM. 
Unfortunately, no
one has yet constructed a phenomenologically successful extension that
has fewer parameters than the SM itself!\footnote{Unless one starts with the
default extension to the SM featuring nonzero neutrino masses, namely the see-saw
model with three right-handed neutrinos.  There are successful models employing
leptonic family symmetries that have fewer parameters \cite{Low}.} 
The basic problem is that the
larger the symmetry of the Lagrangian, the more SSB is required.
The resulting large Higgs sectors of almost all extended models are necessarily full of
arbitrary parameters, with the Higgs potentials often too
complicated to properly analyse. This motivates the search for alternative
SSB mechanisms. The purpose of this paper is to study spontaneous symmetry
breaking in trinification-style grand unified theories (GUTs) \cite{Glashow, Babu,Willenbrock, Carone}.

A new mechanism called the ``clash of symmetries'' \cite{Davidson,Shin} was recently proposed within
the brane-world context \cite{branes} 
(for a general review of brane-world models, see \cite{Rubakov, Csaki} and 
references therein). It still uses elementary scalar fields, but
relaxes the requirement that the background Higgs fields correspond to
the spatially homogeneous configurations given by a single global minimum of
the Higgs potential. Instead, the stable configurations corresponding
to topological kinks or domain walls are used, where the 3-brane
constituting our universe is located at the coincident centres of the
walls. (In fact, it is well-known that branes can be dynamically
generated through the interactions between scalar fields and
higher-dimensional gravity endowed with a cosmological term \cite{extradim}.) 
The spontaneous
breakdown pattern now becomes a function of the extra spatial
coordinate(s), simply because the Higgs configurations vary in this(these)
direction(s). In models with both continuous gauge symmetries and
discrete symmetries, rich patterns of kink-induced SSB are possible,
offering model-building opportunities unavailable in the
standard spatially homogeneous case. If spontaneously broken discrete
symmetries do not feature in a model of interest, then sufficiently
long-lived metastable non-topological kinks are a viable alternative.

One motivation for this paper is to see if this mechanism can
help in the construction of GUTs of
trinification type. These models have gauge group
$SU(3)^3$ and a discrete symmetry permuting the isomorphic
$SU(3)$ factors. Another motivation is simply the
study of kink solutions in such theories for their own sake, a topic
of general interest in field theory, cosmology and condensed matter
physics \cite{Rajaraman}.

The present work complements existing clash of symmetries studies in a
natural way. The mechanism itself was independently discovered by
Davidson, Toner, Volkas and Wali \cite{Davidson} 
with the brane-world application as
motivation, and by Pogosian and Vachaspati \cite{Pogosian} 
in the context of $3+1$-dimensional GUT
theories. Davidson et al constructed a toy model that essentially
consisted of three $SU(3)$ Higgs triplets with a discrete symmetry
under permutations, while Pogosian and Vachaspati focused on adjoint
kinks in $SU(N)$ models, especially $SU(5)$. Here we will study Higgs
fields in $(\mathbf{3,\overline{3}})$ representations, thus naturally
extending the Davidson et al scenario. It is also related in an
interesting way to adjoint kinks, because, under the diagonal subgroup,
$(\mathbf{3,\overline{3}})$ decomposes into a (complex) adjoint and
singlet. Furthermore, the Davidson et al model was a distillation of
ideas fermenting around the GUT group $E_6$, which has the
trinification GUT group $SU(3)^3$ as a subgroup \cite{E6}. The other GUT
subgroup of $E_6$, namely $SO(10)$, has already been partially analysed
from the clash of symmetries perspective \cite{Shin}, and some quite encouraging SSB
outcomes were obtained. One reason for embarking on the present study was to see if the
alternative trinification route from an underlying $E_6$ might also
lead to promising model building opportunities.

In the next section, we review trinification and the clash of symmetries idea.  
Following that, we show that the one-Higgs-multiplet model cannot be made to
work even with kink-induced symmetry breaking.  In Sec.\ \ref{cha:twofields}
we outline a novel way to avoid the intra-multiplet VEV hierarchies required
in standard trinification models.  We conclude in Sec.\ \ref{cha:conclusion}.

\section{Trinification, symmetry breaking and the clash of symmetries}
\label{cha:review}

\subsection{Trinification and standard symmetry breaking}\label{cha:triintro}

The trinification gauge group is
\begin{equation}
G_3 = SU(3)_c \otimes SU(3)_L \otimes SU(3)_R,\label{eq:trigauge}
\end{equation}
where $c$ stands for colour, and $L(R)$ for left(right). One family
of quarks, leptons and exotic partner fermions $\Psi_L$ is placed 
in the anomaly-free, reducible 27-dimensional representation
\begin{equation}
\Psi_L \sim \mathbf{\underline{27}} = 
\left(\mathbf{3,\overline{3},1}\right)+
\left(\mathbf{1,3,\overline{3}}\right)+
\left(\mathbf{\overline{3},1,3}\right).\label{eq:fermions}
\end{equation}
Gauge coupling constant unification is achieved by imposing a discrete
$Z_3$ symmetry that permutes the $SU(3)$ factors.

The physical identification of the components in the multiplets depends
on the identification of the electric charge, equivalently hypercharge, 
generator. It is interesting
to note that there are three possible choices, that is, there are three
embeddings of electric charge and hypercharge
in the trinification gauge group. The same
is also true in $E_6$, as has been commented on several times in the
literature (note that $E_6$ and $G_3$ have the same rank) \cite{E6}.

An interesting feature of trinification (and $E_6$) models is that the
Higgs multiplets $\Phi$ coupling to the fermion bilinears transform the same way as the
fermions,
\begin{equation}
\Phi \sim \left(\mathbf{3,\overline{3},1}\right)+
\left(\mathbf{1,3,\overline{3}}\right)+
\left(\mathbf{\overline{3},1,3}\right),
\end{equation}
making two independent invariant Yukawa terms of 
the form $\overline{(\Psi_L)^c} \Psi_L \Phi$. Previous work \cite{Glashow, Babu}
has shown that at least two copies of such a Higgs multiplet are needed to
achieve realistic spontaneous symmetry breakdown and fermion mass generation
in the context of ordinary $3+1$ dimensional gauge theory.
More recently, Willenbrock \cite{Willenbrock} 
has observed that gauge coupling constant
unification will happen in non-supersymmetric trinification models if
a sufficient $\Phi$ multiplicity is introduced. This is an interesting
observation, even though the resulting model apparently must suffer from
the gauge hierarchy problem. We shall return to the trinification gauge 
hierarchy issue in a later
section. For the moment, let us focus on the two-$\Phi$
scenario, studied in some depth in Ref.\cite{Babu}.

Let the two Higgs multiplets be called $\Phi$ and $\chi$, and denote
the irreducible components as per
\begin{equation} 
\phi_c \sim \left(\mathbf{1,3,\overline{3}}\right),\quad
\phi_L \sim \left(\mathbf{\overline{3},1,3}\right),\quad
{\rm and}\quad
\phi_R \sim \left(\mathbf{3,\overline{3},1}\right),
\end{equation}
and similarly for $\chi$. 
Each irreducible piece is best represented by a $3 \times 3$ matrix.
The transformation law for the colourless field $\phi_c$ is then $\phi_c \to U_L\, \phi_c U_R^{\dagger}$
where $U_{L,R}$ are $SU(3)_{L,R}$ matrices, and so on. We shall denote the generators
of $SU(3)_L$ by $L_{1,\ldots,8}$, similarly for the right-handed group. 

A VEV of the form
\begin{equation}
\langle \phi_c \rangle = \left( \begin{array}{ccc}
0 & 0 & 0 \\ 0 & 0 & 0 \\ 0 & 0 & v
\end{array} \right)
\label{eq:v3}
\end{equation}
spontaneously breaks $SU(3)_L \otimes SU(3)_R $ to a standard left-right symmetric
subgroup $SU(2)_L^{(1,2)} \otimes SU(2)_R^{(1,2)} \otimes U(1)_{B-L}$, where
\begin{equation}
B-L = L_8 + R_8.\label{eq:bl}
\end{equation}
The $SU(2)$ factors act on the top-left $2 \times 2$ block: the superscript $(1,2)$
means `acting on the first and second rows and columns' for the left- and right-handed
groups respectively.

Suppose the second multiplet develops a VEV of the form
\begin{equation}
\langle \chi_c \rangle = \left( \begin{array}{ccc}
0 & 0 & 0 \\ 0 & 0 & 0 \\ 0 & v' & 0
\end{array} \right).
\label{eq:v2}
\end{equation}
The breakdown induced by this VEV on its own is also to a left-right
symmetric subgroup, namely $SU(2)_L^{(1,2)} \otimes SU(2)_R^{(1,3)} \otimes
U(1)_{Y''}$. The right-handed $SU(2)$ is now differently
embedded within the parent $SU(3)_R$, and the
$Y''$ generator is
\begin{equation}
Y'' = L_8 - 2 R_8,
\end{equation}
the third embedding of hypercharge.
The effect of the two VEVs acting together is to break $SU(3)_L \otimes SU(3)_R$
to the {\it intersection} of the two, differently-embedded left-right symmetric subgroups
defined above. The intersection group is precisely the electroweak group
$SU(2)_L^{(1,2)} \otimes U(1)_Y$, where
\begin{equation}
Y = R_3 + L_8 + R_8.
\label{eq:Y}
\end{equation}
It is interesting to note that intersections of differently embedded but isomorphic 
subgroups is also central to the clash of symmetries mechanism.

The other two hypercharge embeddings are obtained by permuting the nonzero
entries along the bottom row of the VEV matrices: $Y'= L_8-R_3+R_8$ requires
them to be in the first and third slots, while $Y''$ needs the first and second slots.

To complete our discussion, we address
electroweak symmetry breaking. For definiteness, we focus on the hypercharge
embedding of Eq.\ (\ref{eq:Y}). Using the definition of
electric charge
\begin{equation}
Q = \frac{1}{2} \left( L_3 + Y \right),
\end{equation}
we see that, up to gauge transformations, the most general VEV patterns inducing complete
breakdown to $U(1)_{Q}$ are
\begin{equation}
\langle \phi_c \rangle =
\left( \begin{array}{ccc}
u_1 & 0 & 0 \\ 0 & u_2 & 0 \\ 0 & 0 & v_1 \end{array} \right),
\qquad 
\langle \chi_c \rangle =
\left( \begin{array}{ccc}
u_3 & 0 & 0 \\ 0 & u_4 & u_5 \\ 0 & v_2 & v_3 \end{array} \right),\label{eq:VEV}
\end{equation}
where the $u$'s are of order the electroweak scale ($\ll v_{1-3}$).
Reference \cite{Babu} explains why having just one multiplet, whose VEV we
can take to be of the form $\langle \phi_c \rangle$ above without
loss of generality, leads to unrealistic fermion masses and mixings.
Unfortunately, the most general quartic potential with both $\Phi$
and $\chi$ has very many independent terms and parameters. This goes
against the fundamental reason for introducing larger gauge symmetries:
the hope of greater predictivity. Furthermore, one requires a delicate
VEV hierarchy {\it within each multiplet}, $u_{1-5} \ll v_{1-3}$. Needless
to say, this necessitates unnatural fine-tuning.

\subsection{Kink-induced symmetry breaking}\label{cha:clash}

Topologically stable or metastable kinks serve as dynamically-induced branes,
and they in general feature symmetry breaking patterns or hierarchies that
depend on the extra dimension coordinate $w$.  

When a symmetry group $G$ spontaneously breaks to a subgroup $H$, the vacuum manifold
is usually given by the coset space $G/H$.  All elements of $G/H$ yield isomorphic
stability groups, but different elements imply different embeddings of the subgroup $H$
within $G$.  When the parent group has a discrete group factor that is spontaneously
broken, the vacuum manifold is disconnected.  It contains a discrete number of
copies of $G/H$ (where $G$ is the continuous part of the parent group).

By definition, kinks are static one-dimensional solutions of the classical Euler-Lagrange
equations that have as boundary conditions that the configurations asymptote to elements
of the vacuum manifold as $w \to +\infty$ and $w \to -\infty$.  If these two elements
live on disconnected pieces, then the kink is topologically stable.  If not, it is
non-topological but possibly metastable or perturbatively stable.  This depends on
the Higgs potential topography:  for perturbative stability, there must be an initial 
energy cost when perturbations to the kink are added.

The simple $Z_2$ kink used in pedagogical discussions arises in a model where the
vacuum manifold is simply two (disconnected) points.  The features we wish to explore
occur when $G/H$ is a non-trivial manifold, because now there are many choices for the
kink boundary conditions.

The pure clash of symmetries mechanism arises when the kink begins and ends on
disconnected pieces of the vacuum manifold such that the stability groups at $w=+\infty$
and $w=-\infty$ are differently embedded (though isomorphic) $H$ subgroups of $G$.  We shall
call these groups $H(+\infty)$ and $H(-\infty)$.
If, by
contrast, they are identically embedded [$H(+\infty) = H(-\infty) \equiv H$], 
then the symmetry group at most points $w$
is just the very same $H$.  With different embeddings asymptotically, the kink has
to perform a non-trivial interpolation.  This means that, in general, the unbroken
symmetry at points $|w|<\infty$ is the intersection $H(+\infty) \cap H(-\infty)$,
so it is smaller than $H$.  The possible advantage here is greater
symmetry breaking power from a given Higgs model compared to the standard case where a
spatially homogeneous VEV configuration is used.  Note that there may be special
points, most commonly $w=0$, where the configuration instantaneously passes 
through a special breaking pattern.  This arises if some of the components
of a multiplet are odd functions of $w$ (such as the hyperbolic tangent),
because at $w=0$ those components vanish and symmetry breaking is reduced.

Higgs multiplets in models with non-trivial $G/H$ manifolds are multi-component.
When we say ``kink configuration'' in this context, we do not mean that all
nonzero components have to be odd functions like the standard $Z_2$ kink.
The components may be odd functions, or even functions, or neither, and these
may coexist within the one multiplet.  It depends on the specific VEVs
to which the configuration asymptotes.  Consider a given component $\phi_i$ of a
multiplet $\Phi$.  An odd function for that component arises 
if $\phi_i(+\infty) = -\phi_i(-\infty)$,
an even function if $\phi_i(+\infty) = +\phi_i(-\infty)$, and neither if
$\phi_i(-\infty) \neq 0$ but $\phi_i(+\infty)=0$ (or vice-versa).  This has
an interesting consequence.  Even if $H(+\infty)=H(-\infty)=H$ so that the
symmetry breaking is almost always to the same $H$, relative sign changes
in the asymptotic components of $\Phi$ can mean that intra-multiplet hierarchies
change as a function of $w$.  In particular, we can hope to exploit the
differences between odd- and even-function components to generate a large hierarchy
in effective VEVs near the brane at $w=0$.  We will use this possibility
in Sec.\ \ref{cha:twofields}.

\section{Symmetry breaking and kinks in $SU(3) \otimes SU(3)$ with
one $(\mathbf{3,\overline{3}})$ Higgs multiplet}\label{cha:onefield}

Symmetry breaking in trinification models is greatly concerned with
VEV patterns for Higgs fields in $(\mathbf{3,\overline{3}})$
representations of an $SU(3)_1 \otimes SU(3)_2$, identified with the extended left-right
group in those models. This provides good motivation for studying Higgs
potentials, symmetry breaking patterns and kink solutions for a
model with a single Higgs field $\Phi \sim (\mathbf{3,\overline{3}})$.
This is also of general interest because it is 
a rich theoretical laboratory despite its apparent
simplicity. It ties in well with studies of $SU(N)$ adjoint Higgs models
since, as already mentioned, $\Phi$ branches into a complex octet and
singlet under the diagonal $SU(3)$ subgroup.

The field $\Phi$ is represented as a general $3 \times 3$ matrix of complex
field components,
\begin{eqnarray}
\centering \Phi=\left( \phi_i^j \right), \qquad \left( \phi_i^j
\right) \in \mathfrak{C}, \qquad i,j=1,2,3,
\end{eqnarray}
undergoing the transformation
\begin{eqnarray}
\centering \Phi \rightarrow U_1 \Phi U_2^{\dagger},
\end{eqnarray}
where $U_{1,2}$ are elements of the fundamental representations
of $SU(3)_{1,2}$, respectively.

\subsection{The quartic potential model}
We studied a model described by a 
quartic $\Phi$ potential in the most detail as quartics
are the simplest non-trivial potentials
which can be bounded from below, and they are renormalisable 
in $3+1$ dimensions. (For now we ignore the question of 
ultraviolet completion in higher dimensional models, leaving it as a 
problem for the future).

The most general quartic $SU(3)_1 \otimes SU(3)_2$ invariant potential is
\begin{eqnarray}
\centering V= -m^2 Tr[\Phi^\dagger \Phi]+ \lambda_1
\left(Tr[\Phi^\dagger \Phi] \right)^2+\lambda_2 Tr[\Phi^\dagger \Phi
\Phi^\dagger \Phi]+(a \: \Phi_i^l \Phi_j^m \Phi_k^n \epsilon^{ijk}
\epsilon_{lmn} + H.c)\label{eq:potential1},
\end{eqnarray}
which is bounded from below in the parameter region
\begin{eqnarray}
\centering \lambda_1 > 0,\qquad \;\textrm{and} \qquad 3\lambda_1 +
\lambda_2 > 0.\label{eq:positivity}
\end{eqnarray}
If $a=0$,
then the potential has an extra $U(1)$ symmetry $\Phi \to e^{i\theta}\Phi$.
With $a \neq 0$, this $U(1)$ is explicitly broken to the discrete $Z_3$
subgroup.  

For homogeneous field configurations, an $SU(3)_1 \otimes SU(3)_2$
transformation can always be used to bring $\Phi$ into diagonal form,
\begin{eqnarray}
\centering \Phi= e^{i\alpha} \left( \begin{array}{ccc} \phi_1 & 0 & 0 \\ 0 &
\phi_2 & 0 \\ 0 & 0 & \phi_3 \end{array} \right),\label{eq:higgs}
\end{eqnarray}
where $\alpha, \phi_i \in R$.
The potential now possesses invariance under a permutation symmetry $S_3$ also. 

Global minimisation revealed four non-trivial extrema that are global minima
 for certain ranges of the Higgs parameters, where these regions of parameter 
space were studied extensively. Although this analysis is important in the 
study of kinks, we do not show it here as it was found that the one-Higgs-field 
model showed no promise in model-building outside the context of the 
pure study of kinks. Rather, we indulge in a brief summary of our results. 

The symmetry breaking patterns instigated by the four global minima leave 
the unbroken subgroups $SU(2)_1 \otimes SU(2)_2 \otimes U(1),\;
SU(3)_{1+2}, \; \textrm{and} \; SU(2)_{1+2} \otimes U(1).$
Clearly, a single Higgs field in conventional trinification models can not
provide sufficient symmetry breaking capabilities unless we invoke 
kink-induced symmetry breaking. 

Invariance of our potential under the discrete symmetries $S_3, Z_3$ means that 
each VEV is actually one choice out of a set of vacua which
yield degenerate global minima. 
These represent our coset spaces $ G/H $ 
which form our vacuum manifold.
Thus, we can obtain VEV configurations which leave differently embedded isomorphic 
subgroups unbroken, and choose these to be the boundary conditions of our 
kinks. At non-asymptotic points, the kink will then respect a reduced symmetry 
given by the intersection of the two groups.

Given the degeneracy of our vacua and the wealth of their parameter space
dependence, there are a large number of distinct kinks which can 
be constructed. In section \ref{cha:clash} it was
outlined how the stability of these domain wall solutions depends on the 
discrete symmetry $Z$ of the theory. 
Topologically non-trivial kinks asymptote
to vacua which are elements of different connected sectors, each being $G/H$, 
while the spontaneously broken group elements of $Z$ map
each sector into one another. As a result of this broken discrete symmetry,
there is an energy barrier between the two vacua, ensuring the stability of 
the kink. However, in this model, the question of 
stability is not so explicitly
defined. The discrete transformations of our quartic potential, $Z_3, \, S_3$, 
are contained in $SU(3)_1 \otimes SU(3)_2$ and so cannot assume the privileges of 
$Z$. Instead, the kinks will only lie in disconnected coset spaces if they
cannot be connected by any continuous symmetry of our Higgs potential. That is,
for any elements $U_1 \in SU(3)_1$ and $U_2 \in SU(3)_2$,  
\begin{eqnarray}
\centering
\langle \Phi \rangle_1 \nrightarrow   \, U_1  
\langle \Phi \rangle_2 U_2^{\dagger}.\label{eq:stability}
\end{eqnarray}
For the special case of $a=0$, the Higgs potential has an additional $U(1)$
symmetry also via which the vacua should not be related. If any two VEV
configurations can be connected by any one of these continuous transformations
then they lie in the same sector $G/H$. Kinks in this instance are 
non-topological, though they might be metastable and sufficiently long-lived
to be useful.

Unfortunately, for all possible cases, it is trivial to find either a 
$U_{1,2}$ such 
that the equality of Eq.\ (\ref{eq:stability}) holds, or a $U(1)$ connecting 
the VEVs. So, although it has a rich family of non-topological configurations, the 
quartic potential one-Higgs-field model does not admit any topological kinks.

We conclude this section by noting that these 
configurations are interesting in their own right, but they are not of
direct applicability for the trinification symmetry breaking problem: almost 
all
of the unbroken $SU(2)$ factors are diagonal
subgroups, leaving only vector-like combinations unbroken with 
the identification of $(1,2)$ with $(L,R)$. 

\subsection{Beyond the quartic potential}

We now study what additional possibilities open up if we consider non-quartic
potentials.
In order for a spatially varying 
Higgs field to be viable, we must associate ourselves with a brane-world 
hypothesis, where the Higgs field propagates in the bulk. 
This complicates the question of ultraviolet completeness. Lacking much 
phenomenological guidance on its structure, we are {\it{a priori}} free to choose any bounded
Higgs potential. A sextic potential for example,
 would contain
terms proportional to (det$\Phi)^2$ which break the $U(1)$ to a $Z_6$ 
symmetry which is not a subgroup of $SU(3) \otimes SU(3)$. Thus, if this $Z_6$
 symmetry was spontaneously broken then the stability of our kinks would be
 guaranteed. So it is possible to find topological kinks for a one field model
for non-quartic Higgs potentials. 

It may also be fairly straightforward to construct non-quartic Higgs potentials such that
non-topological kinks become at least perturbatively stable.  As part of work in progress,
D.P. George and one us (RRV) \cite{Damien} 
has demonstrated that sextic potentials with a sufficiently
deep local (not global) minimum at $\phi=0$ will produce perturbatively
stable kinks interpolating between $\phi=v$ and $\phi= -v$, where $v$ is the
value of $\phi$ at the global minimum. The toy model studied here was of a single
complex scalar field $\phi$ with a $U(1)$ invariant sextic potential. We expect
that this result can be extended to non-Abelian multi-component Higgs models,
though this has not as yet been rigorously checked.

Now, it is not necessary to write down a specific potential in order 
to determine the qualitative pattern of symmetry breaking as a function
of the bulk coordinate. To determine the VEVs that can serve as boundary conditions,
independent of the potential, we need only to write down all possible
diagonal configurations for our single Higgs multiplet, as per Eq.\ (\ref{eq:higgs}). 
This yields the unbroken subgroup $U(1) \otimes U(1)'$ as a possible symmetry breaking 
avenue in addition to those 
of the quartic potential case. 

Consider now any general Higgs potential which
is $SU(3)^2$ invariant and assume that it also has a discrete symmetry which,
firstly, is not contained in $SU(3)^2$, and secondly, is not a subgroup of any additional 
accidental continuous symmetries. 
We wish to find all possible pairs of VEVs which are transformed into
each other by discrete transformations outside $SU(3)^2$. They are
\begin{eqnarray}
\centering
\Phi(-\infty) = \langle \Phi \rangle_7  = diag \left( 0,a , b  \right), 
\qquad 
\Phi(+\infty) = \langle \Phi \rangle_8  = diag \left( 
0, -a, b  \right), \;\;\; + \;\; \textrm{cyc. perms},\label{eq:boundc3} 
\end{eqnarray}
\begin{eqnarray}
\centering
\Phi(-\infty) = \langle \Phi \rangle_9  = diag \left( 
a, b, c \right), \qquad  
\Phi(+\infty)= \left\{
\begin{array}{ll} \langle \Phi \rangle_{10} & = 
diag \left( -a , b, c \right),\;\;\; + \;\; \textrm{cyc. perms} \\ 
\\
\langle \Phi \rangle_{11} & = 
diag \left( -a , -b ,-c \right),\label{eq:boundc4}  
\end{array}\right.
\end{eqnarray}
where $a$, $b$ and $c$ are not necessarily unequal.

We now give an example of the kind of topological kink that can arise
in a non-quartic model.
The most interesting case has kinks interpolating between 
the VEVs of Eq.\ (\ref{eq:boundc3}). Let $\Phi(-\infty)=\langle \Phi \rangle_7$, 
and $\Phi(+\infty)=\langle \Phi \rangle_8$, then
$\Phi(w)=diag \left( 0 , -a\,f(w) , b \right)$,
where $f(w)$ is some kink-like odd function (perhaps a hyperbolic tangent)
whose specific form of course depends on the Higgs potential chosen.
Independent of this choice, we see that the $w$-dependent symmetry breaking
pattern has
$H(-\infty)=H(+\infty)=H_{clash}=U(1) \otimes U(1)'$, 
but on the brane, at $w=0$, $\Phi(0)=diag \left( 0 , 0,b 
 \right)$ and the symmetry is enhanced to
$H(0)=SU(2)_1 \otimes SU(2)_2 \otimes U(1)$. 
We see that while the left-right symmetry group is exact at $w=0$, the additional breakdown
off the brane is quite strong, to $U(1) \otimes U(1)'$. 

We conclude that a rich family of topologically stable kinks may be produced by 
models with non-quartic Higgs potentials. This is interesting, though from
the point of view of symmetry breaking in trinification models none of the
additional configurations possible for non-quartic potentials
seem to be of direct use, as they generate the wrong breaking patterns.

\section{Trinification with two $(\mathbf{3,\overline{3}})$ Higgs multiplets}
\label{cha:twofields}

\subsection{Application to the trinification hierarchy problem}

We have seen that the one-field model is an interesting theoretical laboratory
in terms of VEV patterns, associated kinks (either topological or not
depending on the case) and symmetry breaking patterns. However we did not find
any case that was obviously promising for the trinification application,
even when kink-induced symmetry breaking was used. We thus now 
turn to the two field case.

Reiterating the introductory discussion, the trinification gauge group is
given by
\begin{eqnarray}
\centering G_3=SU(3)_c \otimes SU(3)_L \otimes SU(3)_R,
\end{eqnarray}
which is augmented by the cyclic symmetry $Z_3$ to ensure a single
gauge coupling constant. 
The symmetry breaking to the SM gauge group is compactly accomplished
with the VEV configuration of Eq.\ (\ref{eq:VEV})
where $u \sim 10^2 GeV$ and $v \sim 10^{15} GeV$. The gauge
hierarchy and fine-tuning problems in this model are quite delicate as there are
hierarchies in the VEVs within a single multiplet, there being
entries of both the unification and
electroweak breaking scales.
We will see that the introduction of the spatial variation in the Higgs field
may be able to alleviate the hierarchy problem in a novel way. We restrict ourselves to 
considering only two colour-singlet Higgs fields $\phi_c,\, \chi_c$ 
which transform under Eq.\ (\ref{eq:trigauge}) 
as $\phi_c$, $\chi_c \sim (1,\mathbf{3,\overline{3}})$.

The two colour-singlet Higgs 
fields are allowed  
to have the most general spatial variation in the extra dimensional
coordinate,
\begin{eqnarray}
\centering
\phi_c (w) = \left( \begin{array}{ccc}
\varphi_1 (w) & 0 & 0 \\
0 & \varphi_2 (w)  & 0 \\ 
0 & 0 & \varphi_3 (w) 
\end{array} \right), \qquad
\chi_c (w) = \left( \begin{array}{ccc}
\chi_1 (w) & 0 & 0 \\
0 & \chi_2 (w) & \chi_3 (w)\\
0 & \chi_4 (w) & \chi_5 (w)
\end{array} \right), \label{eq:spatial}
\end{eqnarray}
but we relax the constraint that they leave differently embedded, isomorphic 
subgroups unbroken at
$|w| = \infty$ as would be demanded if we were considering the ``pure'' clash idea. 
The VEVs retain the structural form of Eq.\ (\ref{eq:VEV}), however all 
entries are elevated to be of the unification order. With this modification, 
the model now admits only one SSB scale, eliminating 
the hierarchy. We will see that a hierarchy in observed symmetry breaking
scales may nevertheless be effectively generated through the details of the
spatial variation of the various components within $\phi_c$ and $\chi_c$.

At the asymptotic points $|w| =\infty$, the Higgs fields must approach 
VEV profiles 
which are related to each other via a discrete symmetry of the potential.
We do not wish to specify a Higgs potential, but rather complete the most 
generalised analysis possible. So we consider all the VEV configurations which
may be related by a discrete symmetry of some potential. The most general 
set of discrete transformations are independent reflection 
symmetries in each 
individual field component. 
Requiring these VEVs to be
disconnected from each other greatly restricts $\phi_c$,
giving only four possible choices. If $\phi(-\infty) \equiv \phi_v^1 = $
$diag \left( u_1, u_2, v_1 \right)$,
then 
\begin{eqnarray}
\centering
\phi_c (+\infty) &\equiv& \phi_v^2 = diag \left( 
-u_1, u_2 , v_1 \right), \qquad \;\;\;\;\;\;
\phi_c (+\infty) \equiv \phi_v^3 = diag \left( 
u_1 , -u_2 , v_1 \right), \nonumber
\\ \nonumber
\phi_c (+\infty) &\equiv& \phi_v^4 = diag \left( 
u_1 , u_2 , -v_1 \right) \;\;\;\; \textrm{and} \;\;\;\; \;
\phi_c (+\infty) \equiv \phi_v^5 = diag \left( -u_1, -u_2 , -v_1 
\right). \nonumber
\end{eqnarray}
The field $\chi_c$ is much less restricted. There are a large number of distinct 
combinations of VEVs which are not connected by an $SU(3)^3$ transformation. 
We can narrow this set by observing the behaviour of the field on the brane.
The profile of the field on the brane is determined by considering 
the parity of the configurations which interpolate between these VEVs. 
That is, we can ascertain 
which components pass through zero on the brane without the explicit 
derivation of the form of the kinks. If the $\chi_1^1$ component is non-zero 
on the brane, then the $SU(2)_L^{(1,2)}$ subgroup will be broken. However, 
this is exactly the weak isospin gauge group of the Weinberg-Salam model which 
must remain unbroken for realistic model-building. So we need only 
consider the VEVs which respect $\chi_1(-\infty) = -\chi_1(+\infty)$, 
reducing the number of choices. 
Of these, there are only 20 distinct configurations on the brane, and 
given that it is the symmetry breaking on the brane with which 
we are concerned, we can focus our analysis on these 
without loss of generality.

Together, with the possibilities for $\phi_c$, there are 80 unique possible kink configurations 
for a model described by the trinification gauge group.
Consider 
first the symmetry breaking initiated by  $\phi_c$. If  
$\phi_c(-\infty) = \phi_v^1$ and $\phi_c(+\infty) = \phi_v^2, \;\phi_v^3$ or 
$\phi_v^4$, then $\phi_c(0)$ has two non-zero entries on the brane.
The best case scenario for these choices will give a maximum 
unbroken subgroup of 
$SU(2)_L \otimes SU(2)_R \stackrel{\phi_c(0)}{\rightarrow} SU(2)_{L+R} \otimes U(1)$. Hence, 
when the intersection is taken with the appropriate $\chi$ profile, the 
largest unbroken symmetry possible on the brane is  
$SU(3)_c \otimes SU(2)_{L+R} \otimes U(1)$, subject to
specific parameter constraints. All other instances have the chiral 
$SU(3)$'s break to $U(1)$'s. 
This is clearly insufficient in the model-building perspective because weak 
isospin is vector-like. 

As a result, there is a unique choice for $\phi_c(w)$, given by 
$\phi(-\infty)=\phi_v^1$ and $\phi(+\infty)=\phi_v^5$, leaving only 20 
distinct configurations on the brane to consider. As 
$\phi_v^5=-\phi_v^1$, all of the components of $\phi_c(w)$ pass through 
zero at $w=0$, leaving 
$\chi(0)$ solely responsible for all the breaking on the brane.
Analysing the $\chi_c(w)$ configurations, there is a unique pattern of symmetry 
breaking on the brane that offers merit. This is given by the boundary 
conditions 
\begin{eqnarray}
\centering
\chi_v^{1} \equiv \chi_c (- \infty) = 
\left(\begin{array}{ccc} 
u_3 & 0 & 0 \\ 
0 & u_4 & 0 \\ 
0 & v_2 & v_3 \end{array} \right), \qquad 
\chi_v^{2} \equiv \chi_c (+\infty) =
\left( \begin{array}{ccc} 
-u_3 & 0 & 0 \\ 
0 & -u_4 & 0 \\ 
0 & v_2 & -v_3 \end{array} \right),\label{eq:tribc1}
\end{eqnarray} 
where we have performed an $SU(2)_R$ transformation on $\chi_c$ to rotate 
$u_5$ to zero. Despite the rotation, these boundary conditions are still 
disconnected via an $SU(3)^3$ transformation, with the rotation chosen 
in order to obtain the 
correctly embedded $SU(2)_L$ subgroup.

On the brane, 
\begin{eqnarray}
\centering
\chi(0) = \left( \begin{array}{ccc}
0 & 0 & 0 \\
0 & 0 & 0 \\
0 & v_2 & 0 \end{array} \right)
\end{eqnarray}
which instigates the breaking 
\begin{eqnarray}
\centering SU(3)_c \otimes SU(3)_L \otimes SU(3)_R \rightarrow SU(3)_c
\otimes SU(2)_L \otimes SU(2)_R \otimes U(1)_{Y''}.\label{eq:leftrt}
\end{eqnarray}
Enforcing the brane world hypothesis realistically to this model may
provide a novel way to understand the gauge hierarchy as first explored
in \cite{Shin}. The assignment
of $w$ as the extra dimensional coordinate, enables the
Higgs configuration to propagate through the extra dimensions. The
brane is located at $w=0$, on which the SM degrees of freedom are
confined. Actually the degrees of freedom for the brane world theory
are larger than the SM. The theory localised to the brane is the full
modified trinification theory described above, its Lagrangian
described by the symmetry of $G_3$. The brane world
fields then interact with the Higgs fields $\phi_c,\, \chi_c$, that
instigate symmetry breaking.

Recall that
our VEV entries are all of the unification scale, so as the fields vary
spatially in $w$, their components range from $0-10^{15} GeV$ in a natural way. 
On the brane, $\phi_c (0) = 0$ and $\chi_c (0)$ has one non-zero 
component $v_2 \sim 10^{15} GeV$, and this Higgs field \textit{strongly} 
induces the breakdown of Eq.\ (\ref{eq:leftrt}) to a 
left-right symmetric group. It is difficult to believe that our trinification 
theory has a definite, sharp localisation at $w=0$. If the
brane world degrees of freedom are permitted a slight leakage off the
wall, then the effective brane world theory will also be affected by
the Higgs field for small, finite values of $w$ centred about the
brane. This instigates an additional symmetry breakdown, but at a
much \textit{weaker} scale, the strength being proportional to the amount
of leakage, and equivalently, the distance scale over which the Higgs field 
components substantially vary. At these points, the Higgs configurations are of the form
of Eq.\ (\ref{eq:spatial}). Individually, each field breaks the left-right
symmetry, the intersection of their unbroken subgroups determining the
symmetry in the bulk. $\phi_v^{5}$ generates the breaking
\begin{eqnarray}
\centering 
SU(3)_c \otimes SU(2)_L \otimes SU(2)_R \otimes U(1)_{Y''}
\rightarrow SU(3)_c \otimes U(1)_{L_3+R_3} \otimes U(1)_{L_8+R_8},
\end{eqnarray}
while $\chi_v^{1}$ is responsible for the breaking
\begin{eqnarray}
\centering 
SU(3)_c \otimes SU(2)_L \otimes SU(2)_R \otimes U(1)_{Y''}
\rightarrow SU(3)_c \otimes U(1) \otimes U(1)'.
\end{eqnarray} 
Together, the two multiplets induce the breaking down to the colour
and QED gauge groups
\begin{eqnarray}
\centering 
SU(3)_c \otimes SU(2)_L \otimes SU(2)_R \otimes U(1)_{Y''}
\rightarrow SU(3)_c \otimes U(1)_{Q}.\label{eq:lfrtqed}
\end{eqnarray} 
So we have `naturally' achieved electroweak breaking without having to 
introduce additional, unnatural VEV components in the Higgs fields, 
and/or additional Higgs fields.  
In this context, the gauge hierarchy issue which plagues SSB can be
attributed to the structural form of the kink solutions, with emphasis
on their spatial variation, and the characteristics of the leakage off
the brane. It is the amount of leakage and the value of the Higgs
field at this finite $w$ which determines the scale of the electroweak
breaking. If a well-motivated brane localisation mechanism can be
found that allows exponentially small leakage, then the gauge
hierarchy would be explained.

With the exception of the degrees of freedom which enjoy the leakage,
if the observable 3+1 dimensional universe is confined to this brane,
then the symmetry group is not the SM, but a left-right symmetric
theory. Although Eq.\ (\ref{eq:lfrtqed}) represents a necessary
relationship for the model to be compatible with observed electroweak
breaking, there is the need for the pattern
\begin{eqnarray}
\centering 
SU(3)_c \otimes SU(2)_L \otimes SU(2)_R \otimes
U(1)_{Y''}\rightarrow SU(3)_c \otimes SU(2)_L \otimes U(1)_{Y}
\label{eq:lfrtsmm}
\end{eqnarray}
to be an intermediary. (The hypercharge generator of the SM group here has to 
be the conventional embedding so as to be consistent with the breaking of 
Eq.\ (\ref{eq:lfrtqed})).  
The SM is an established effective field theory at low energies and we must 
be able to generate its gauge group along this symmetry breaking avenue.
It is not sufficient to have only the left-right symmetric theory of 
Eq.\ (\ref{eq:leftrt}) on the brane, with the breaking of the chiral $SU(2)_R$
and electroweak symmetry occurring at the same stage, and hence same energy, in the bulk. 
Left-right symmetric theories, realised with a $U(1)_{B-L}$ factor, 
have been extensively studied, with the breaking 
$SU(3)_c \otimes SU(2)_L \otimes SU(2)_R \otimes
U(1)_{B-L}\rightarrow SU(3)_c \otimes SU(2)_L \otimes U(1)_{Y}$ 
currently having a phenomenological lower bound of $\sim 10^{3}$ GeV,
whereas the electroweak 
breaking occurs at $M_{EW}\sim 10^2 GeV$. 
Therefore the two symmetry breaking stages
\begin{eqnarray}
\centering
SU(3)_c \otimes SU(2)_L \otimes SU(2)_R \otimes
U(1)_{Y''}& \rightarrow & SU(3)_c \otimes SU(2)_L \otimes U(1)_{Y} \\
 SU(3)_c \otimes SU(2)_L \otimes U(1)_{Y} & \rightarrow & SU(3)_c \otimes U(1)_Q
\end{eqnarray} 
may represent a difference in scale by a
factor of no more than ten.
The form of the kink
configurations provides a possible solution. If $\varphi_3(w)$ has a slightly
steeper gradient as the $\phi_c$ Higgs profile approaches the brane, then, to
leading order, Eq.\ (\ref{eq:lfrtsmm}) would occur. The factor of ten may have its 
origins in a small spread in the Higgs
parameters. Subsequently, the total symmetry breaking pattern produced
may be
\begin{eqnarray}
\centering 
SU(3)_c \otimes SU(3)_L \otimes SU(3)_R \rightarrow 
SU(3)_c \otimes SU(2)_L \otimes SU(2)_R \otimes U(1)_{Y''} \rightarrow 
G_{SM}
\rightarrow SU(3)_c \otimes U(1)_{Q}.
\end{eqnarray}
This two Higgs model based upon trinification theory, therefore, has 
shown the potential for kink profiles that generate a 
phenomenologically acceptable symmetry breaking
pattern. Furthermore, these kinks and the breaking for 
which they are responsible, have been constructed without reference to any 
specific Higgs potential. This means that the question of the existence 
and stability of our solution reduces to finding an appropriate potential.

As the above shows, the brane-world kink method holds promise for inducing 
the correct symmetry breaking 
patterns necessary for trinification models. However, our analysis
is obviously incomplete.  The main unresolved issues are the
incorporation of gravity and the localisation of fields to the brane.
The former is likely to be relatively straightforward, but the latter
presents some important challenges.
The leakage scenario itself is plausible, and inevitable in
the case of dynamical localisation where wave-functions are
sharply peaked around the brane but not delta-function-like. 
For the case of one extra dimension, it is known that several ingredients
are needed to achieve dynamical localisation of fermions, Higgs fields (if that is
necessary) and gauge fields \cite{localisation}.
To conclusively demonstrate that the ideas presented above, which we
think are interesting and novel, can be the basis for a realistic brane-world
model requires considerably more work and is beyond the scope of
this paper.

\section{conclusion}
\label{cha:conclusion}

Motivated by the attractive trinification route to grand unification, we have
examined symmetry breaking patterns and kink solutions in models with
one or two Higgs multiplets assigned to the $(\mathbf{3,\overline{3}})$
representation of an $SU(3)_1 \otimes SU(3)_2$ gauge symmetry.

For the one field case, a rich pattern of symmetry breaking outcomes
was rigorously deduced for the case of a quartic Higgs potential. Employing
the resulting global minima as boundary conditions, we studied
the possible kink configurations of the model.  We found that there are many possible
non-topological kink configurations, but none that are topologically stable.
One motivation for this analysis was to see
if the clash of symmetries mechanism of Refs.\ \cite{Davidson}
and \cite{Shin} and generalisations thereof 
might be of use in a brane-world realisation of trinification,
with the Higgs field propagating in the bulk. Unfortunately, none of the
bulk-coordinate-dependent symmetry breaking patterns identified appear to
be useful in this regard.  We discussed how topologically stable kinks
can arise when non-quartic potentials are used, recognising that the
restriction to quartic form is not mandatory for brane-world models with
their unknown ultraviolet completion.  The
strategy here was to identify those discrete transforms of Higgs global
minima configurations that are outside the continuous $SU(3)_1 \otimes SU(3)_2$ gauge
group, and to use the spontaneous breaking of those discrete invariances
to ensure topological stability.  The assumption here was that one could
construct a non-quartic Higgs potential so as to realise the discrete symmetries
required. 

We then turned to the two-field case in the context of the brane-world trinification
application.  In the usual $3+1$-dimension incarnation of the model, two
$(\mathbf{3,\overline{3}})$ Higgs multiplets are required to ensure phenomenologically
acceptable symmetry breaking and fermion mass generation. However there is a
severe fine-tuning problem: some components of each multiplet must be given unification
scale values, while others must be of electroweak scale. We put forward
what we think is an interesting, albeit speculative, alternative: Go to
a brane-world realisation with one extra dimension, and have {\it all} nonzero components
of the Higgs fields acquire unification-scale values only.  Then effectively
induce the electroweak to unification scale hierarchy by exploiting the spatial
dependence of the Higgs components with respect to the bulk coordinate. Those
components that are odd functions (say of hyperbolic tangent form) go to
zero exactly on the brane, despite the fact that they asymptote to
unification scale values infinitely far from the brane. By exploiting these
zeros and postulating mild differences in the slopes of the kinked
components near the brane, we found a situation where a hierarchical 
cascade, trinification to left-right symmetry to the standard model to
just colour cross electromagnetism, might ensue.

\acknowledgments{This work was supported in part by the Australian
Research Council and in part by the Commonwealth of Australia.}

\end{document}